\RequirePackage{ifpdf}
\documentclass[letterpaper]{JHEP3}
\pdfoutput=1
\usepackage{graphicx}
\usepackage{amsmath}
\usepackage{amsfonts}
\usepackage{epsfig}
\usepackage{slashed}
\usepackage{braket}

\usepackage{comment}

\def\bec{\begin{center}}
\def\eec{\end{center}}
\def\beq{\begin{equation}}
\def\eeq{\end{equation}}
\def\bea{\begin{eqnarray}}
\def\eea{\end{eqnarray}}

\def\KD{K\"{a}hler-Dirac }
\def\Omegab{{\overline{\Omega}}}
\def\omegab{{\overline{\omega}}}

\def\partialb{\overline{\partial}}
\def\etab{\overline{\eta}}
\title{Topological fermion condensates from anomalies}

\def\psib{\overline{\psi}}

\def\Phib{\overline{\Phi}}

\author{Simon Catterall, Jack Laiho and Judah Unmuth-Yockey\\
Department of Physics, Syracuse University, Syracuse, NY 13244, USA \\
E-mail: \email{smcatter@syr.edu}
}

\abstract{We show that a class of fermion theory formulated on a compact, curved manifold will generate a 
condensate whose magnitude is determined only by the volume and Euler characteristic of the space.
The construction requires that the fermions be treated as K\"{a}hler-Dirac fields and the condensate arises from an anomaly associated with
a $U(1)$ global symmetry which is subsequently broken to a discrete subgroup.
Remarkably the anomaly survives under discretization of the space which allows us to
compute the condensate on an arbitrary triangulation.  The results, being topological in character, should hold
in a wide range of gravitationally coupled fermion theories both classical and quantum.}
\begin{document}
\section{Introduction}
\KD fermions have been proposed as an alternative to Dirac fermions in \cite{Banks:1982iq} and play an important
role in certain supersymmetric field theories where they have been used to construct lattice theories
with exact supersymmetry \cite{Catterall:2009it}. 
They also bear a close connection to staggered fermions on regular hypercubic
lattices \cite{Banks:1982iq,Susskind:1976jm}. 

In this paper we will show that \KD fermions possess another important property; on compact even dimensional
spaces with non-zero Euler
character they exhibit an anomaly which breaks an exact $U(1)$ symmetry and generates a non-zero
fermion condensate.

Rather remarkably this anomaly can be reproduced exactly in a discrete version of the theory, a result
which is tied to the structure of the \KD operator and homology theory. We use this fact to compute the
anomaly and condensate directly on an arbitrary triangulation.

\section{K\"{a}hler-Dirac fermions in the continuum}
\KD fermions naturally
arise on taking the square root of the Laplacian operator written in the language of exterior derivatives.
Consider the anti-hermitian operator
$K=d-d^\dagger$ with $d^\dagger$ the adjoint of $d$ and $d^2=\left(d^\dagger\right)^2=0$.
Clearly $K^2=-dd^\dagger-d^\dagger d=\Box$.
Following Dirac a natural equation for fermions is then given by
\beq
(K-m)\Omega=0\label{KDeq}\eeq
where $\Omega=\left(\omega_0,\omega_1,\ldots \omega_p,\omega_D\right)$ is a collection of $p$-forms (antisymmetric tensors). 
This is the K\"{a}hler-Dirac equation and was first proposed as an equation describing fermions in \cite{Kahler}.
The action of $d$ and $d^\dagger$ on these component forms is given by \cite{Banks:1982iq}
\beq
d\Omega=\left(0,\partial_\mu\omega, \partial_{\mu_1}\omega_{\mu_2}-\partial_{\mu_2}\omega_{\mu_1},\ldots ,\sum_{perms\, \pi}\left(-1\right)^\pi\partial_{\mu_1}\omega_{\mu_2\ldots\mu_D}\right)\eeq
\beq
-d^\dagger\Omega=\left(\omega^\nu,\omega^\nu_\mu,\ldots ,\omega^\nu_{\mu_1,\ldots ,\mu_{D-1}},0\right)_{;\nu}.\eeq
An inner product of two such \KD fields is then given by
\beq
\left[A,B\right]=\int d^Dx\,\sqrt{g}\sum_p \frac{1}{p!}a^{{\mu_1}\ldots{\mu_D}}b_{{\mu_1}\ldots{\mu_D}}.\eeq
Equation~\ref{KDeq} can be obtained by variation of the action 
\beq
S_{\rm KD}=\int d^Dx\sqrt{g}\,\left[\Omegab \left( K-m\right)\Omega\right]\eeq
where $\Omegab$ is an independent (in Euclidean space) \KD field.
From the \KD field $\Omega$ one can build a matrix $\Psi$ using the Clifford algebra of gamma matrices
\beq
\Psi=\sum_{p=0}^D\frac{1}{p!} \gamma^{{\mu_1}\ldots{\mu_p}}\omega_{{\mu_1}\ldots{\mu_p}}.
\label{KDtomatrix}\eeq
In flat space it is straightforward to show that this fermionic matrix satisfies the usual Dirac equation 
\beq
\left(\gamma^\mu\partial_\mu -m\right)\Psi=0\eeq
and describes $2^{D/2}$ degenerate Dirac spinors. Remarkably,
eqn.~\ref{KDeq} holds on any smooth manifold so that the coupling of \KD fields to gravity is somewhat
different from that of Dirac fermions. This difference can be made manifest if one
introduces vielbeins  $\gamma_\mu=e_\mu^a\gamma^a$ in eqn.~\ref{KDtomatrix} 
and replaces the derivative in eqn.~\ref{KDeq} with an appropriate covariant derivative
containing the spin connection $\omega_\mu^{ab}$. This yields a equation which couples the four
spinors encountered in flat space through gravitational interactions. It is easily seen that
this is {\it not} (four copies of) the usual Dirac equation in curved space.  In this paper we will exploit the fact that \KD fermions do not require the introduction of this extra spinor structure and work directly with
eqn.~\ref{KDeq}. 

Consider the linear operator $\Gamma$ which transforms the $p$-form fields according to
\beq\Gamma:\quad \omega_p\to \left(-1\right)^p\omega_p.\eeq  It is straightforward to verify that this operator anticommutes with the
\KD operator $\{K,\Gamma\}=0$. This in turn allows us to define a $U(1)$ symmetry of the massless \KD action
\begin{eqnarray}
\Omega&\to& e^{i\theta\Gamma}\Omega\\\nonumber
\Omegab&\to& \Omegab e^{i\theta\Gamma}.
\end{eqnarray}
This symmetry, which is the analog of chiral symmetry for the Dirac theory,
ensures that any perturbative mass renormalization of the theory is multiplicative in the bare fermion
mass on any background space. It also guarantees that non-zero eigenvalues of $K$ come in pairs, $\pm \lambda_n$,
with corresponding eigenvectors $\phi_n$ and $\Gamma\phi_n$ where
\beq
K\phi_n=\lambda_n\phi_n.
\eeq
Because $K$ is anti-Hermitian, the eigenvalues are purely imaginary.
Of course the symmetry will be broken or anomalous 
if the fermion measure is not invariant under $\Gamma$. Performing such a transformation yields a Jacobian
$J={\rm det}\left(e^{2i\Gamma \theta}\right)$. Thus the existence of the anomaly depends on a non-vanishing trace for $\Gamma$
\beq
{\rm Tr}\,\Gamma=\sum_n \bra{\phi_n}\Gamma\ket{\phi_n}=n_+-n_-\eeq
where the states with non zero eigenvalue are orthogonal and do not contribute while
$n_+$ and $n_-$ denote the number of zero modes with eigenvalue $\pm 1$ respectively\footnote{$\Gamma$ and $K$ commute on the $\lambda=0$
subspace and hence zero modes can be labeled with a given eigenvalue of $\Gamma$.}.
All of this discussion mirrors the usual arguments given for chiral fermions\footnote{In flat space
it is easy to see that $\Gamma$ acts like an axial rotation on the spinors corresponding to columns of
the matrix $\Psi$.} and a derivation of the anomaly
can be gotten in the continuum after the theory is suitably regulated. However, in this case a simpler
method is available. The \KD operator can be discretized using a triangulation of the space in such
a way as to preserve the anomaly exactly. 

\section{\KD fermions on the lattice}

A triangulation of a $D$-dimensional manifold is constructed by gluing together $D$-simplices.
A $p$-simplex consists of $(p+1)$ vertices with every vertex connected to every other vertex. We will
restrict our discussion to triangulations in which each $D$-simplex is equilateral with edge length $a$.
Such triangulations appear in so-called {\it dynamical triangulation} models of quantum gravity \cite{Ambjorn:2010rx,Laiho:2016nlp}. It is
straightforward to generalize this discussion to triangulations with variable edge lengths.

A $p$-simplex contains
$(p+1)$ boundary components which are $(p-1)$-simplices.  The simplest triangulations or simplicial
complexes are constructed by gluing $D$-simplices together along their faces with each face common to two
and only two $D$-simplices. The possible vertex labelings of a given $D$-simplex fall into two categories according to whether they consist of an odd or even permutation of the ordered list of vertices. These two categories are
called orientations.  To define the K\"{a}hler-Dirac operator requires that the triangulation be {\it oriented} which
means that the orientation of each $D$-simplex be chosen in such a way that a given face is held with opposite
orientations in the two $D$-simplices that contain it.

There is a natural mapping of differential forms on a continuum manifold to such an
{\it oriented} discrete triangulation.
The idea is to associate a
continuum $p$-form with a $p$-cochain which is a real or complex linear function over $p$-simplices, $f(C_p)$,
where an individual
$p$-simplex is denoted $C_p$ \cite{Rabin:1981qj,Becher:1982ud}.
The operators $d$ and $d^\dagger$ have corresponding discrete analogs 
in terms of so-called co-boundary, $\partialb$,  and boundary, $\partial$, operators that act on such $p$-simplices. For example the boundary operator
$\partial$ acts on a $p$-simplex consisting of the (ordered) vertices $\left[a_0,\ldots ,a_p\right]$ as
\beq
\partial \left[a_0,\ldots ,a_p\right]=\sum_i\left(-1\right)^i\left[a_0,\ldots ,\hat{a}_i,\ldots, a_p\right]\eeq
where the hat on vertex $\hat{a}_i$ indicates
that the vertex is omitted. Thus the operation of
$\partial$ on some $p$-simplex yields an oriented sum of
$(p-1)$-simplices in its boundary. 
\beq
\partial f(C_{p})=\sum_{C_{p-1}} I(C_{p}, C_{p-1})f(C_{p-1}),\eeq
where $I(C_p,C_{p-1})$ is the incidence matrix for the triangulation which takes the value $+1$ if $C_{p-1}$ is contained
in the boundary of $C_p$ with the correct orientation, $-1$ if it is contained in the boundary with the opposite
orientation, and is zero otherwise.
Similarly the co-boundary operator $\partialb$ is defined by
\beq
\partialb f(C_{p})=\sum_{C_{p+1}}I^T(C_{p}, C_{p+1})f(C_{p+1}).\eeq
From these definitions one can verify that $\partial^2=\partialb^2=0$.
The discrete K\"{a}hler-Dirac operator is defined by
\beq
K_D=\partial-\partialb\eeq
which squares to the discrete Laplacian. Furthermore it should
be clear that $\{\Gamma,K_D\}=0$ when $\Gamma$ acts on $p$-cochains instead of $p$-forms and therefore there is still an exact $U(1)$ symmetry present just as for the continuum theory, acting now
on $p$-cochains as
\beq
\omega_p\to e^{i\theta\left(-1\right)^p}\omega_p\label{Gamsym}.
\eeq
For any given triangulation the fermion measure is just
\beq
\prod_{i_0=1}^{N_0} d\omega_{i_0} d\omegab_{i_0}\prod_{i_1=1}^{N_1}d\omega_{i_1} d\omegab_{i_1}
\cdots\prod_{i_D=1}^{N_D}d\omega_{{i_D}}d\omegab_{{i_D}}\label{measure}
\eeq
where $N_p$ is the number of $p$-simplices in the triangulation and the indices $i_p$ run from
$1\ldots N_p$.  The $\omega_{i_{p}}$ is the fermion field associated with the i\textsuperscript{th} $p$-simplex.
Under the $U(1)$ symmetry in eqn.~\ref{Gamsym} this measure transforms by the phase factor $e^{i2\theta\chi}$ where 
the Euler character $\chi$ is given by \cite{me}
\beq
\chi=N_0-N_1+\ldots+\left(-1\right)^pN_p.\eeq
If $\chi\ne 0$ this breaks the $U(1)$ down to $Z_{2\chi}$, which depends only on the Euler characteristic and is otherwise
independent of
the triangulation. 

Since the non-zero eigenvalues of the lattice operator $K_D$ again come in complex conjugate pairs this phase must
arise from the zero modes of the discrete \KD operator, which must again be either even or odd under $\Gamma$.
Homology theory guarantees that such exact zero modes exist even in the discrete system and
are simultaneously zero modes of the discrete Laplacian  
\beq
\Box=-\partial\partialb-\partialb\partial=-\sum_{C_{p-1}} I(C_{p}, C_{p-1})I^T(C_{p-1}, C_{p}) -
\sum_{C_{p+1}} I^T(C_{p}, C_{p+1})I(C_{p+1}, C_{p})\label{laplacian}.
\eeq
By the Hodge theorem (which applies equally to both discrete
and continuum theories) we know that $n_+ - n_-=\chi$, which is consistent with the previous computation
of the phase change of the
measure. Notice that the anomaly vanishes in odd dimensions since $\chi=0$ there and $n_+=n_-$.

\section{Condensates}
From this point we confine our discussion to the sphere $S^D$ with $\chi=2$ although it is trivial to
generalize to arbitrary Euler characteristic.
In this case  there are precisely 2 zero modes of the Laplacian associated
with the $0$ and $D$-forms (co-chains) both of which have eigenvalues of $\Gamma$ equal to one.

Let us now consider the partition function on a discrete
triangulation $T$ of the sphere
with $N_D$ $D$-simplices and $N_0$ vertices in the presence of fermionic sources $\eta,\etab$ of \KD type, paying careful attention to
the would-be zero modes that arise in the massless limit. 
\beq
Z\left(\eta,\etab\right)={\rm det}\left(K_D^\prime(T)+m\right)e^{\left[\etab\left(K_D^\prime(T)+m\right)^{-1}\eta\right]}\times m^2e^{\sum_{i=1}^2\frac{1}{m}\etab_0^i\eta_0^i}\eeq
where the prime denotes that zero modes satisfying $\Box\Phi^i_0=0$ have been omitted and 
\beq
\eta_0^i=\left[\eta \Phib_0^i\right]\qquad \etab_0^i=\left[\etab \Phi_0^i\right]\eeq corresponds to the ${\rm i^{th}}$ zero mode component of the source.
Notice that the first non-zero contribution to $Z$ as $m\to 0$
occurs by expanding the zero mode sources to quadratic order and yields
\beq
Z\left(\eta,\etab\right)={\rm det}\left(K_D^\prime(T)+m\right)e^{\left[\etab\left(K_D^\prime(T)+m\right)^{-1}\eta\right]}
\etab_0^1\eta_0^1\etab_0^2\eta_0^2.\eeq
The structure of this expression resembles the usual 't Hooft vertex seen in QCD \cite{tHooft:1976rip} and generates a non-local
four fermion interaction in the theory.
Differentiating with respect to the sources will yield a non-zero four fermion correlation function that survives the $m=0$ limit
and breaks the $U(1)$ symmetry down to $Z_4$.
\beq
\left\langle \Omegab(x)\Omega(y)\Omegab(z)\Omega(w) \right\rangle=\frac{1}{Z}\frac{\delta}{\delta\eta(x)}\frac{\delta}{\delta \etab(y)}
\frac{\delta}{\delta\eta(z)}\frac{\delta}{\delta\etab(w)}Z\label{4fermi}\eeq
where $x,y,z,w$ range over the discrete set of coordinates on the triangulation.
To compute this correlator we consider a triangulation in which all simplices are equilateral with fixed
side length.
The discrete Laplacian is block diagonal in the $p$-forms. The $0$-simplex block matrix follows from eq.~\ref{laplacian} and takes the form
\beq
-\Box_{ij}=q_i\delta_{ij}-C_{ij}\label{lap}\eeq
where $C_{ij}=1$ if the vertex $j$ is neighbor to vertex $i$ and $q_i$ is the total number of neighbor vertices.
The first zero mode solution then corresponds to a vector with constant non-zero 0-form components only
\beq
\Phi_0^p=A\delta^{p0}\left(1,1,\ldots\right).\eeq
Normalizing this vector to unity yields $A=1/\sqrt{N_0}$.
The second zero mode is gotten from the $D$-simplex block of the Laplacian and has a structure similar to that given
in eqn~\ref{lap} with the matrix element $C_{ij}$ non-zero for each face separating simplices and 
$q_i=(D+1)$ since each $D$-simplex possesses $(D+1)$ neighbors. The corresponding zero mode
is 
\beq
\Phi_0^p=\frac{1}{\sqrt{N_D}}\delta^{pD}\left(1,1,\ldots\right).\eeq
Inserting these results into eqn.~\ref{4fermi} shows that the correlator in eqn~\ref{4fermi} becomes
\beq \langle \omegab_0(x)\omega_0(y)\omegab_D(z)\omega_D(w) \rangle =\frac{1}{N_0N_D}.\label{4pt}\eeq
Using Wick's theorem this can be written as the product of two-point functions
\beq
\left\langle \omegab_0(x)\omega_0(y)\right\rangle\left\langle \omegab_D(z)\omega_D(w)\right\rangle\eeq
where correlators like $\langle \omegab_0(x)\omega_D(z) \rangle$ and $\langle \omegab_0(x)\omegab_D(z)\rangle$ vanish since the
inverse fermion operator can be written as $\left(-K+m\right)/\left(-\Box+m^2\right)$ which only couples $p$-cochain fields to
to $p$ or $p\pm 1$ co-chains. Furthermore, by translation invariance these two point functions depend only on the difference of
their spacetime arguments and since the RHS of eqn.~\ref{4pt} is independent of coordinates this implies the presence of two
bilinear condensates spontaneously breaking $Z_4$ to
$Z_2$
\begin{eqnarray}
\left\langle \omegab_0(x)\omega_0(x) \right\rangle&=&\frac{1}{N_0}\\\nonumber
\left\langle \omegab_D(x)\omega_D(x)\right\rangle&=&\frac{1}{N_D}
\end{eqnarray}
These lattice expressions
can be expressed in physical units by writing $N_0$ and $N_D$ in terms of the physical volume $V$ and the cut-off $a$.
For example the $D$-simplex condensate is given by $\frac{1}{a^{D-1}}\left(V/a^D\right)^{-1}$.
Note that the magnitude of this condensate depends only on the volume of the space and the Euler number 
and is independent of the triangulation or continuum metric.

%The presence of this four fermion condensate induces a corresponding fermion mass. For example simple dimensional
%analysis suggests that  the mass of the scalar fermion
%$\omega_0$ should be  given by $m=\frac{1}{a}\left(V/a^D\right)^{-\frac{1}{D-1}}$.
%Again notice that this anomaly induced fermion mass is only dependent on the volume and the topology.

\section{Numerical results}
We can illustrate this effect using results obtained from an ensemble of random triangulations of the two sphere
corresponding to the partition function
\beq
Z=\sum_T e^{-S(T)}\eeq
where the action
\beq
S(T)=-\beta\sum_i^{N_0}\log{q_i} +\kappa_2 N_2\eeq
This action contains both a bare cosmological constant $\kappa_2$ and a coupling $\beta$ that controls the local
connectivity of the triangulation. 

The code we use is described in \cite{me} and generates an ensemble of so-called
combinatorial triangulations in which each simplex is uniquely specified by its vertices. These differ
from the degenerate triangulations employed in, for example, \cite{Laiho:2016nlp} although none of our results depend on
this distinction. The code uses a Metropolis algorithm to sample the space of triangulations using
a set of ergodic local changes to the triangulation described in \cite{me}. We generate ensembles
of $10000$ configurations with each configuration separated by 10 sweeps (the average acceptance rate of the algorithm is 10\%). Errors are assessed by binning the data in the standard way and looking for stability as the bin size is
varied. 

\begin{figure}[h]
\begin{center}
\includegraphics[width=0.75\textwidth]{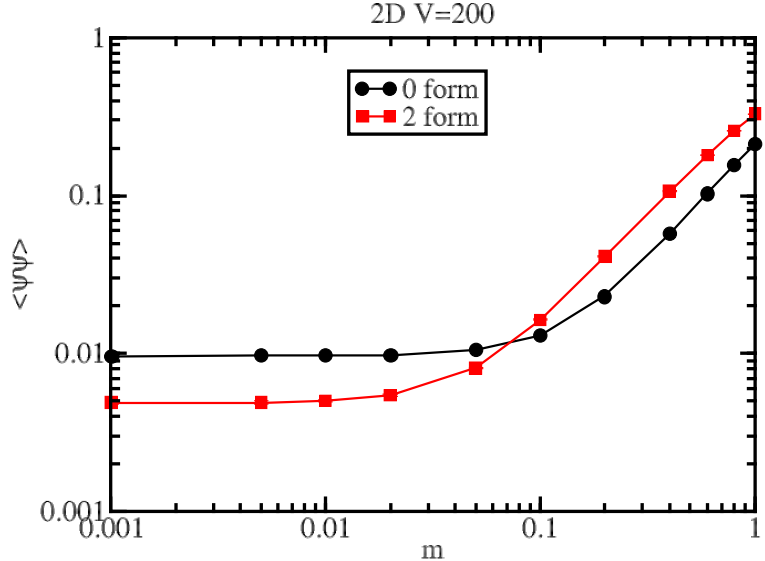}
\caption{Bilinear condensates vs bare mass $m$ on $S^2$ with $N_2=200$ and $\beta=0$\label{pbpvsm}}
\end{center}
\end{figure}

\begin{figure}[h]
\begin{center}
\includegraphics[width=0.75\textwidth]{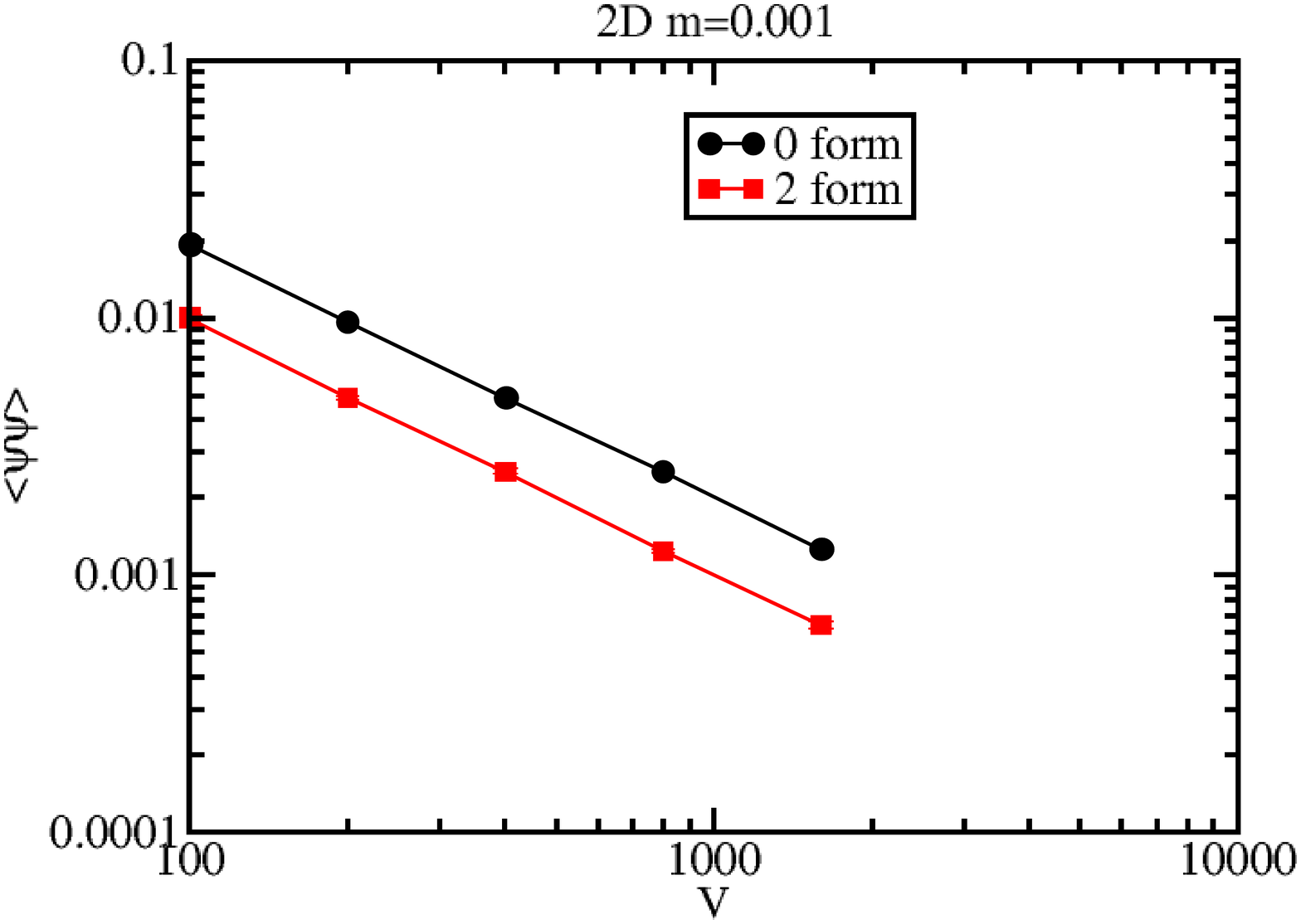}
\caption{Bilinear  condensates  vs volume on $S^2$ with $m=0.001$ and $\beta=0$ \label{pbpvsV}}
\end{center}
\end{figure}

We show in fig.~\ref{pbpvsm} both the $0$ and $2$-form 
bilinear condensates  as a function of the bare mass $m$ (the $1$-form condensate vanishes as expected for $m\to 0$). Notice that the anomaly generated
condensate in the theory with {\it dynamical} \KD fermions is extracted by multiplying the condensate measured in the quenched ensembles by the fermion
mass. Hence  we show $m\left\langle \psib\psi\right\rangle$ in our plots and tables.
We see that the observed value
matches very well with theoretical expectations.
The volume dependence of these condensates at $\beta=0$ and $m=0.001$ is shown in fig.~\ref{pbpvsV} and shows the expected
$1/V$ behavior.

We can further explore the topological character of these condensates by varying $\beta$, which changes the types of triangulation that dominate $Z$ but does not change the global topology.
Table.~\ref{table1} shows the condensates for $\beta=-4.0,0.0,4.0$ at $m=0.001$ and $N_2=200$.
\begin{table}
\begin{center}
\begin{tabular}{||c|c|c|c||}\hline
.			    &$\beta=4.0$	&$\beta=0.0$		&$\beta=-4.0$ 	\\\hline
$m\langle \omegab_0\omega_0 \rangle $&0.0101(2)		&0.0096(2)		&0.0097(2)	\\\hline
$m\langle \omegab_2\omega_2 \rangle $&0.00491(7)		&0.00490(9)		&0.00476(8)	\\\hline
\end{tabular}
\caption{\label{table1} Condensates for several values of the measure coupling $\beta$ for $S^2$ using $m=0.001$
and $N_2=200$, $N_0=102$}
\end{center}
\end{table}

It should be clear that the magnitude of the condensates do not even depend on the dimensionality of the triangulation. Simulations of the four sphere
yield the same condensates taking into account the change in $N_0$ with $\beta$ (in two dimensions $2(N_0-2)=N_2$).
\begin{table}
\begin{center}
\begin{tabular}{||c|c|c|c||}\hline
.			    &$\beta=4.0$	&$\beta=0.0$		&$\beta=-4.0$ 	\\\hline
$m\langle \omegab_0\omega_0 \rangle $&0.0186(3)	&0.034(1)		        &0.053(1)	\\\hline
$m\langle \omegab_4\omega_4 \rangle $&0.0050(1)	&0.0050(1)		&0.0051(1)\\\hline
$1/N_0$                 &0.0187(1)   &0.0341(1)		&0.0527(1)\\\hline
\end{tabular}
\end{center}
\caption{Dependence of condensates on measure coupling $\beta$ for $S^4$ using $N_4=200$ and bare mass $m=0.001$}
\end{table}

\section{Conclusions}
In this paper we have discussed an anomaly that arises in a theory of \KD fermions formulated on a compact
manifold. While anomalies are usually thought of as arising only in continuum theories, in this case the anomaly
survives intact in a discretized version of the theory. 
The reason is simple; the anomaly is sensitive only to the topology of the
background space which can be captured exactly in the lattice theory as a result of the
precise correspondence between
the theory of differential forms and the theory of co-chains - the subject of homology theory.
As in Fujikawa's derivation of the chiral anomaly, this topological anomaly arises from the non-invariance of
the fermion measure under a particular $U(1)$ symmetry specific to \KD fermions.

We have explained in some detail this correspondence and how the discrete \KD operator is
constructed on a triangulation. We should add that our work constitutes the first numerical
study of \KD fermions on random simplicial lattices. In the context of dynamical triangulation models of quantum gravity the 
inclusion of \KD fermions
can potentially play a crucial role in determining both the phase structure of the system and the nature of
the typical geometries that dominate the partition function.  
We hope to examine some of these issues in upcoming work\cite{us}.
Here, however, we would like to stress that the anomaly, being topological in character, is completely agnostic about 
the properties of background space or the nature of any gravitational
fluctuations  -- the effect we describe will be true for a wide range of gravitationally
coupled theories involving \KD fermions in regimes described by both classical and quantum gravity both on
and off the lattice.

It is interesting to ask whether these results could have any implication for cosmology. 
Since \KD fermions behave {\it locally} like multiple copies of Dirac fermions in regions where the curvature is small 
they are not necessarily ruled out by experiment. Of course there are also
global differences between \KD fermions and Dirac fermions. Unlike Dirac fermions, \KD fermions typically possess zero
modes even on positive curvature manifolds which is the origin of the anomaly discussed in this paper.\footnote{This also
makes them interesting from the point of view of Kaluza-Klein reduction since it allows for light fermions to appear in the
compactified theory.} 
The physical manifestation of this anomaly is the existence of a fermion condensate. The magnitude of the latter is
set by the Planck scale times the inverse volume of the Universe in Planckian units. 
Presumably such a condensate would feed into the dynamics of the Universe at early times and might have implications for
cosmology.

\acknowledgments This work is supported in part by the U.S.~Department of Energy, Office of Science, Office of High Energy Physics, under Award Number DE-SC0009998. The authors are grateful for discussions with Jay Hubisz.
\bibliographystyle{JHEP3}
\bibliography{kdanom}

\end{document}